\documentclass[12pt]{article}
\usepackage{ascmac}
\usepackage{latexsym}
\usepackage[dvipdfmx]{graphicx}
\usepackage{amsmath}
\usepackage[top=3.5cm,bottom=3.5cm,left=2cm,right=2cm]{geometry}
\usepackage{pifont}          
\usepackage{amsmath,amssymb}
\usepackage{color}
\usepackage{geometry}
\usepackage{caption}
\usepackage{ulem}
\newcommand{\h}{\hspace}
\newcommand{\be}{\begin{equation}}
\newcommand{\e}{\end{equation}}
\newcommand{\aln}[1]{\begin{align}#1\end{align}}

\begin{document}

\title{
\vbox{
\baselineskip 14pt
\hfill \hbox{\normalsize KUNS-2640
}} \vskip 1cm
\bf \Large  
Possible explanations for fine-tuning of the universe
\vskip 0.5cm
}
\author{
Kiyoharu~Kawana\thanks{E-mail: \tt kiyokawa@gauge.scphys.kyoto-u.ac.jp}
\bigskip\\
%\\*[20pt]
\it \normalsize
 Department of Physics, Kyoto University, Kyoto 606-8502, Japan\\
\smallskip
}
\date{\today}

\maketitle

\abstract{\normalsize
The Froggatt-Nielsen mechanism and the %Coleman's
multi-local field theory are interesting and promising candidates for solving the naturalness problem in the universe. These theories are based on the different physical principles: The former assumes the micro-canonical partition function $\int {\cal{D}}\phi\ \prod_i \delta (S_i^{}-I_i^{})$, and the latter assumes the partition function $\int {\cal{D}}\phi\ \exp\left(iS_M^{}\right)$ where $S_M^{}$ is the multi-local action $\sum_i c_i^{}S_i^{}+\sum_{i,j}c_{i,j}^{}S_i^{}S_j^{}+\cdots $. Our main purpose is to show that they are equivalent in the sense that they predict the same fine-tuning mechanism. In order to clarify our argument, we first study (review) the similarity between the Froggatt-Nielsen mechanism and statistical mechanics in detail, and show that the dynamical fine-tuning in the former picture can be understood completely in the same way as the determination of the temperature in the latter picture. Afterward, we discuss the multi-local theory and the equivalence between it and the the Froggatt-Nielsen mechanism.
Because the multi-local field theory can be obtained from physics at the Planck/String scale, this equivalence indicates that the micro-canonical picture can also originate in such physics. As a concrete example, we also review the IIB matrix model as an origin of the multi-local theory.

%In other words, it is possible to understand that the Froggatt-Nielsen mechanism is just one of the aspects of the Planck/String scale physics.
}
\newpage

Although the Standard Model (SM) is completed by the discovery of the Higgs boson, there are many open questions in it such as the Higgs quadratic divergence, the Strong CP problem, the cosmological constant problem, and so on. These problems are difficult to answer in ordinary quantum field theory (QFT), and called the naturalness problem. Therefore, it is quite important to seek for new theory or mechanism that naturally answers these questions.

One of the possibilities is to try to explain the observed couplings %as the dynamically tuned values by making them dynamical
by dynamical fine-tuning. For example, in the Strong CP problem, $\theta$ becomes dynamical by considering the Peccei-Quinn symmetry and its breaking. However, even if such a field theoretical approach with a new symmetry can solve one of the fine-tunings, it is difficult to solve a few problems simultaneously. %Of course, each of the problems might be rooted in different physics,
So, it is meaningful to study new mechanism that can realize a few fine-tunings simultaneously.
 
Among various proposals, the Froggatt-Niselsen mechanism (FNM) \cite{MPP1} recently attracts much attention because the predicted value of the  Higgs mass ($\sim 130$GeV) was close to the observed value $\simeq 125$GeV. % well known as one of such mechanisms.
It was originally proposed to explain the nontrivial behavior of the SM Higgs potential at high energy scale: The potential has another minimum around the Planck scale, and it can degenerate with the electroweak vacuum $v_h^{}=246$GeV depending on the values of the SM couplings. Such a degeneration is called the Multiple point criticality principle (MPP), and there are a lot of studies so far \cite{MPP2}. The fundamental assumption in the FNM is to use the micro-canonical partition function %even in QFT 
like statistical mechanics, and its origin still remains obscure. In this picture, the couplings in QFT become dynamical, and their dynamical fine-tuning can take place. See \cite{MPP1,MPP2} and the following discussion for the details.

On the other hand, in \cite{Hamada:2015dja}, it was also argued that a few naturalness problems, including the MPP, can be solved by the multi-local field theory. It assumes that the effective action below the Planck/String scale is given by the multi-local one: $\sum_i c_i^{}S_i^{}+\sum_{i,j}c_{i,j}^{}S_i^{}S_j^{}+\cdots$, where $S_i^{}$'s are ordinary actions, and $c_i^{}$'s, $c_{ij}^{}$'s, $\cdots$ are constants. Although it seems difficult to study the theory as quantum theory, we will see that we can reduce it to QFT with the couplings being dynamical by a simple mathematical transformation. See the following discussion for the details. As well as the FNM, we do not need to consider its fundamental origin as long as we apply it to the naturalness problem, %This theory also makes a big assumption that the effective action is given by the multi-local action $\sum_i c_i^{}S_i^{}+\sum_{i,j}c_{i,j}^{}S_i^{}S_j^{}+\cdots$. 
however, such an origin can be actually found in physics at the Planck/String scale. Therefore, the multi-local theory seems to be more promising than the FNM it that everything can be explained from more fundamental physics. %We also give a short review of \cite{Asano:2012mn} where the multi-local action is obtained from the IIB matrix model.

The purpose of this paper is to show that these two approaches are in fact equivalent in the sense that they predict the same fine-tuning mechanism in QFT: The coupling in QFT is fixed at the point that most strongly dominates in their partition functions, and the fine-tuned value generally depends on the details of the theories. This fact indicates that the micro-canonical picture may also originate in the Planck/String scale physics such as the wormhole theory \cite{Coleman:1988tj} or matrix model. As a concrete example, we also review the derivation of the multi-local theory from the IIB matrix model \cite{Asano:2012mn}. Although the study in \cite{Asano:2012mn} is mathematically rigorous, most of the discussion can be done without relying on the details of mathematics. So, in this paper, we aim to give an instructive and intuitive explanation of their work.
\\

\begin{figure}[!t]
\begin{center}
\includegraphics[width=9cm]{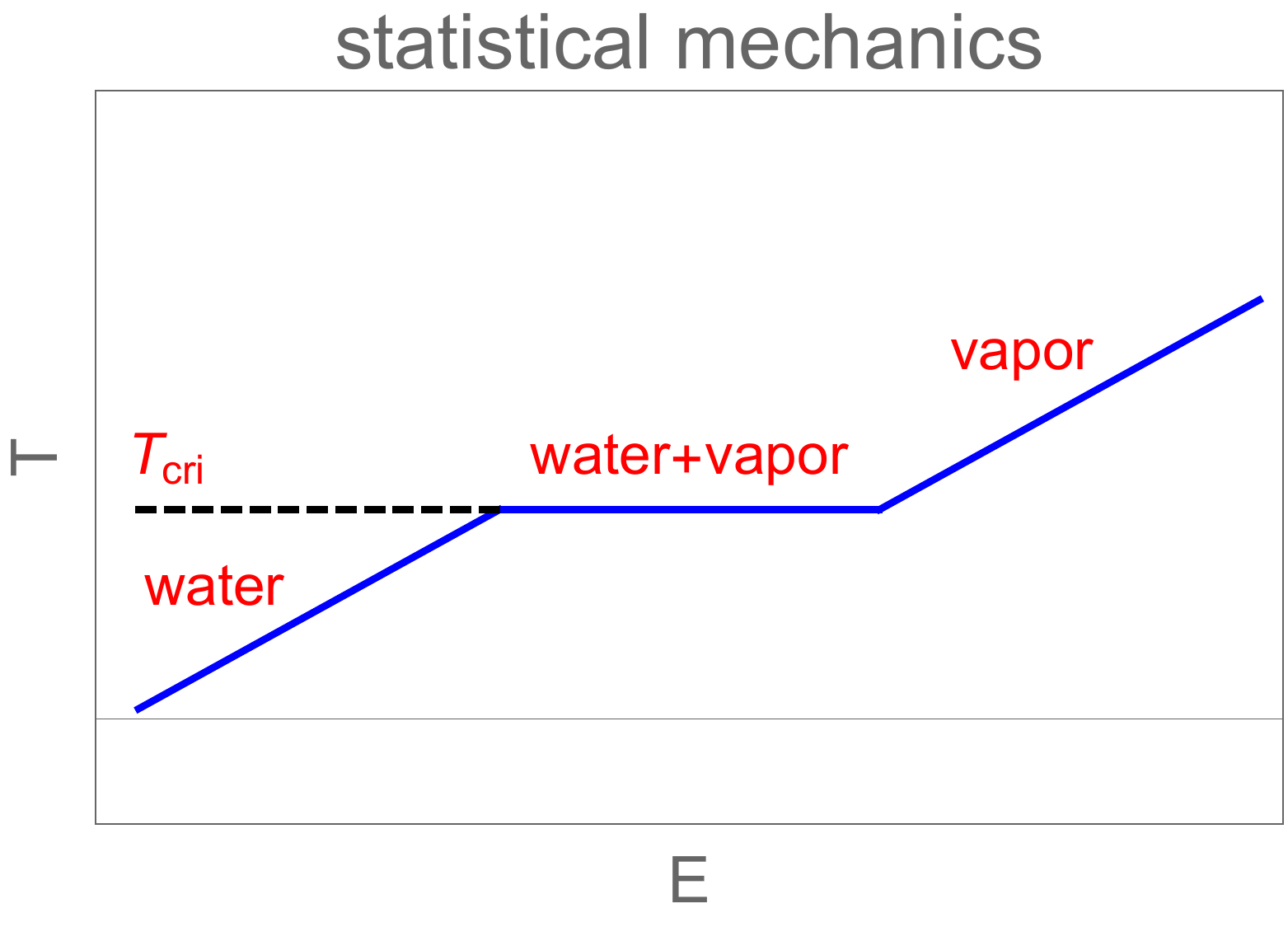}
\end{center}
\caption{The blue contour shows $T^*(E)$ determined by the micro-canonical distribution. For example, consider water heated by external environment. While the phase transition occurs, $T^*(E)$ does not change. 
}
\label{fig:phase1}
\end{figure}

Let us first review the FNM. In ordinary QFT, a system is completely described by the partition function:
\be Z^{\text{(QFT)}}(\lambda) =\int {\cal{D}}\phi\ \exp\left(iS\right),
\e
where $S$ is a given action, and $\lambda$ represents a coupling in $S$. The corresponding quantity in statistical mechanics is the canonical distribution:
\be Z_{\text{C}}^{(\text{Statistical})}(\beta)=\sum_{n} \exp\left(-\beta E_n^{}\right),
\e
where $\beta=1/T$ is the inverse temperature, and $E_n^{}$'s are the energy eigenvalues. On the other hand, it is known that this distribution is equivalent to the micro-canonical distribution
\be Z_{\text{MC}}^{(\text{Statistical})}(E)=i\pi\sum_n\delta(E_n^{}-E)\label{eq:micro canonical}
\e
in the thermodynamic limit
\footnote{The brief proof is as follows: When the space volume $V_3^{}\rightarrow\infty$, the canonical distribution can be written as
\be Z^{(\text{Statistical})}_{\text{C}}=-\beta \int_{0}^\infty dE\Omega(E)e^{-\beta E}=-\beta V_3^{} \int_{0}^\infty d\epsilon \ \exp\left(V_3^{}(s-\beta \epsilon)\right)\sim \beta V_3^{}\exp\left(V_3^{}(s-\beta \epsilon)\right)\bigg|_{\epsilon=\epsilon^*},
\e
where $\Omega(E)$ is the number of states, $S(E)=\log\Omega(E)$ is the entropy, $s$ is its density, $\epsilon$ is the energy density, and $\epsilon^*$ is the solution of $ds/d\epsilon=\beta$. Thus, the free energy is given by
\be F(\beta)=-\frac{1}{\beta}\log Z^{(\text{Statistical})}_{\text{C}}\sim \frac{V_3^{}}{\beta}(\beta\epsilon^*-s(\epsilon^*))=\underset{E}{\text{Min}}\left(E-TS(E)\right)
\e
This shows that the free energy determined by the canonical distribution is thermodynamically equivalent to the entropy defined by the micro-canonical distribution.
}.
Here, we have added the overall coefficient $i\pi$ just for the following argument. From the microscopic point of view, the micro-canonical distribution is more fundamental because there is no thermodynamical quantity in Eq.(\ref{eq:micro canonical}). In this picture, the temperature $T$ is dynamically determined by the following way: Eq.(\ref{eq:micro canonical}) can be rewritten as
\aln{&=\sum_n \left[\frac{1}{E_n^{}-E}-P\left(\frac{1}{E_n^{}-E}\right)\right]\nonumber
\\
&=\int_0^{\infty}d\beta\ \sum_n\exp\left(-\beta(E_n^{}-E)\right) -\sum_nP\left(\frac{1}{E_n^{}-E}\right)\nonumber
\\
&\simeq \int_0^{\infty}d\beta\ Z^{(\text{Statistical})}_{\text{C}}(\beta)e^{\beta E}=\int_0^{\infty}d\beta\ \exp\left(-\beta (F(\beta)-E)\right),
}
where $P$ represents the Cauchy principal value, and $F(\beta)$ is the free energy. In the thermodynamic limit, this integral is dominated by $\beta^*(E)=1/T^*(E)$ that satisfies
\be\frac{\partial }{\partial\beta}\left(\beta F(\beta)\right)\bigg|_{\beta=\beta^*}-E=0\ \leftrightarrow\ \langle H\rangle_{\text{C}}=E,
\e
where $H$ is the Hamiltonian of the system, and $\langle\ \rangle_{\text{C}}$ represents the average by the canonical distribution. Thus, $T$ is dynamically fixed so that the average of $H$ by the canonical distribution can become $E$, and its value depends on $E$. Although E is completely arbitrary from the microscopic point of view, there is a special interval where $T^*(E)$ does not depend on $E$: coexisting of different phases. See Fig.\ref{fig:phase1} for example. This figure shows a schematic contour of $T^*(E)$. When different phases coexist, $T^*(E)=T_{\text{cri}}$ does not change until the system gains enough energy.  %In other words, when a system experiences the phase transition, $T^*$ is given independently of $E$ (or microscopic details). 
From the naturalness point of view, this fact indicates that $T$ is most likely to be fixed at the critical value $T_{\text{cri}}$ because the probability of $E$ being chosen to be in the interval is biggest. In the following, we will apply the above argument to QFT, and see that the coupling in QFT corresponds to $T$ in statistical mechanics. 

%____________________________Phase__________________________________
%_________________________________________________________________________

By considering the above discussion, it is natural to adopt the micro-canonical partition function even in QFT:
\be Z^{\text{(QFT)}}_{\text{MC}}=\int {\cal{D}}\phi\ \prod_i\delta(S_i^{}-I_i^{}),\label{eq:mc QFT}
\e
where $S_i^{}$'s are the ordinary local actions in QFT, and $I_i^{}$'s are their arbitrary values. %For example, if we choose the Higgs mass term
%\be \mu^2\int d^4x\ H^\dagger H,
%\e 
%we expect
%\be I_{H}^{} \sim V_4^{}M_{\text{pl}}^4,
%\e
%where $V_4^{}$ is the spacetime volume. 
In principle, $S_i^{}$'s in Eq.(\ref{eq:mc QFT}) should be determined from the microscopic physics such as String theory \footnote{For now, we assume that all the low-energy (renormalizable) actions in the SM are included in $S_i^{}$'s. %composed of there is no principle that constrains which $S_i^{}$ should appear in Eq.(\ref{eq:mc QFT}). However, from the low-energy or Wilsonian point of view, it is sufficient to consider the renormalizable ones. 
}. Note that $I_i^{}$'s correspond to $E$ in statistical mechanics. Rewriting the delta function in Eq.(\ref{eq:mc QFT}) as the Fourier form, we obtain
\aln{ Z^{\text{(QFT)}}_{\text{MC}}&=\int {\cal{D}}\phi\ \int\cdots\int \left(\prod_id\lambda_i^{}\right)\ \exp\left(i\sum_i\lambda_i^{}(S_i^{}-I_i^{})\right)\nonumber
\\
&=\int\cdots\int \left(\prod_id\lambda_i^{}\right)\ Z^{(\text{QFT})}(\overrightarrow{\lambda})=\int\cdots\int \left(\prod_id\lambda_i^{}\right)\ \exp\left(-iV_4^{}\rho(\overrightarrow{\lambda})\right),\label{eq:mc QFT2}
}
where we have introduced the Lagrange multipliers $\lambda_i^{}$'s, and $\rho(\overrightarrow{\lambda})$ is the vacuum energy density. One can see that $\lambda_i^{}$'s play roles of the couplings. Thus, if there is a point $\overrightarrow{\lambda}_0^{}$ that strongly dominates in the above integration, we have
\be \sim Z^{(\text{QFT})}(\overrightarrow{\lambda}_0^{}),
\e
and this is the ordinary partition function where $\overrightarrow{\lambda}$ is fixed to $\overrightarrow{\lambda}_0^{}$. Here, note that such a dominant point is naively given by the saddle point of $\rho(\overrightarrow{\lambda})$ \footnote{Of course, it is possible that a system can not satisfy Eq.(\ref{eq:saddle point}) for any value of $I_i^{}$. In this case, we must carefully study the coupling dependence of $\rho(\overrightarrow{\lambda})$. See \cite{MLT} for example. When a saddle point exits, the fluctuation of the coupling is roughly given by ${\cal{O}}\left(((\rho''(\lambda)V_4^{})^{-1/2}\right)$.
}:
\be \frac{\partial\rho(\overrightarrow{\lambda})}{\partial\lambda_i^{}}=\frac{\int {\cal{D}}\phi\ (S_i^{}-I_i^{})\exp\left(i\sum_{j}\lambda_j^{}(S_j^{}-I_j^{})\right)}{Z^{(\text{QFT})}(\overrightarrow{\lambda})}=\langle S_i^{}\rangle-I_i^{}=0.\label{eq:saddle point}
\e
In \cite{MPP1}, Froggatt and Nielsen showed that, for the wide range of $I_i^{}$, the above saddle point corresponds to the coexisting phase of the Higgs vacua like the phase transition in statistical mechanics. As one of examples, let us choose the Higgs quartic term $S_H^{}=\int d^4x\ \left(H^\dagger H\right)^2$ as $S_i^{}$, and confirm their claim. Because the Higgs potential can have two minima depending on the couplings in the SM, we denote the small (large) vacuum expectation value as $\phi_1^{}(\lambda)$ ($\phi_2^{}(\lambda)$) in the following discussion. Furthermore, we represent the critical Higgs quartic coupling where the two vacua degenerate as $\lambda_{\text{cri}}^{}$. %and $\phi_1^{}(\lambda_{\text{cri}})$ ($\phi_2^{}(\lambda_{\text{cri}})$) at the point as $\phi_1^{}(\lambda_{\text{cri}})=\phi_{\text{cri}}^{}$ ($$).
 When $I_H^{}\leq \phi_1(\lambda_\text{cri}^{})^{4}V_4^{}\ (V_4^{}:\text{spacetime volume})$, the Higgs quartic coupling $\lambda$ is fixed at the point $\lambda^*(I_H^{})\ (\geq
\lambda_{\text{cri}}^{})$ so that $\phi_1^{}(\lambda)$ can satisfy Eq.(\ref{eq:saddle point}):
\be \langle S_H^{}\rangle=\phi_1^{}(\lambda^*(I_H^{}))^4V_4^{}=I_H^{}.\label{eq:saddle2}
\e

%____________________________Phase QFT__________________________________
\begin{figure}
\begin{center}
\begin{tabular}{c}
\begin{minipage}{0.5\hsize}
\begin{center}
\includegraphics[width=8.5cm]{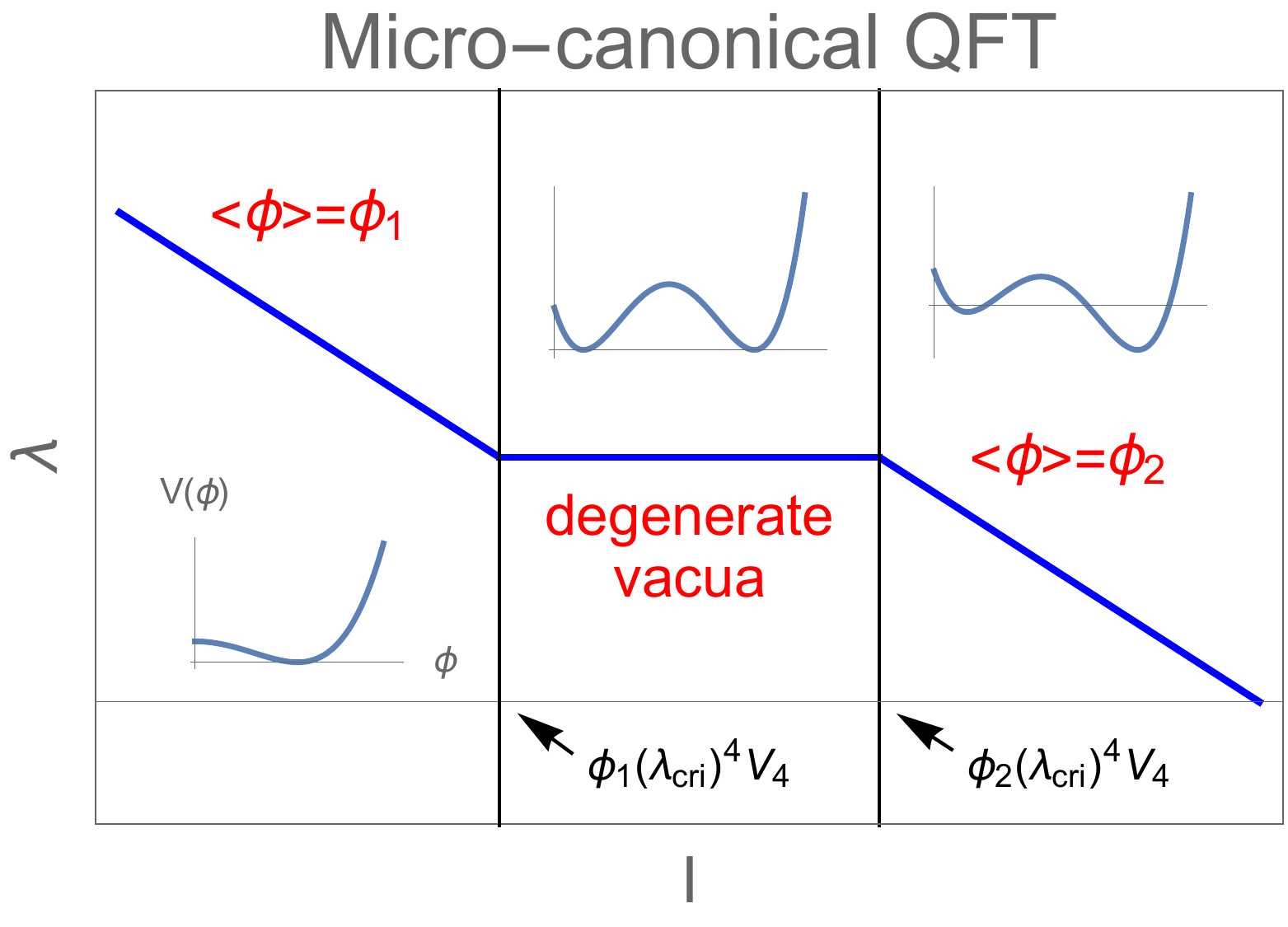}
\end{center}
\end{minipage}
\begin{minipage}{0.5\hsize}
\begin{center}
\includegraphics[width=7.5cm]{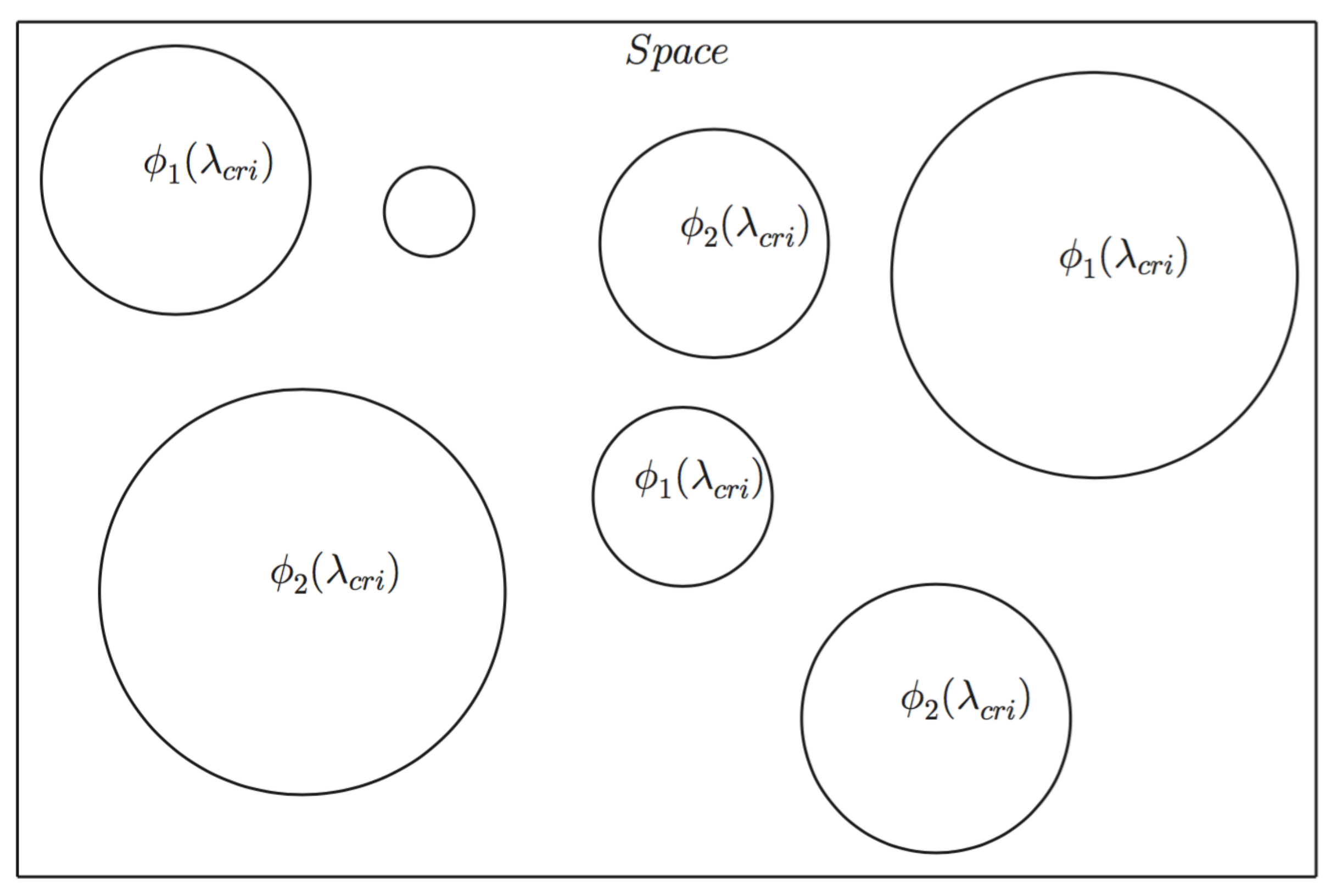}
\end{center}
\end{minipage}
\end{tabular}
\end{center}
\caption{Left: The fixed Higgs quartic coupling $\lambda$ as a function of $I_H^{}$. As well as the statistical mechanics, there is a finite interval $\phi_1^{}(\lambda_{\text{cri}}^{})^4V_4^{}\leq I_H^{}\leq \phi_2^{}(\lambda_{\text{cri}}^{})^4V_4^{}$ where the phase coexisting is realized. Right: The coexisting of the Higgs vacua when $\phi_1^{}(\lambda_{\text{cri}}^{})^4V_4^{}\leq I_H^{}\leq \phi_2^{}(\lambda_{\text{cri}}^{})^4V_4^{}$.
}
\label{fig:phase2}
\end{figure}
%_________________________________________________________________________

\noindent In the left panel in Fig.\ref{fig:phase2}, we show $\lambda^*(I_H^{})$ by a blue contour. Note that $\phi_1^{}(\lambda)$ is the true vacuum in the this case. As we increase $I_H^{}$, $\lambda^*(I_H^{})$ decreases in order to satisfy Eq.(\ref{eq:saddle2}). When $I_H^{}$ becomes $\phi_1^{}(\lambda_{\text{cri}})^4V_4^{}$, namely, $\lambda^{*}(I_H^{})$ becomes $\lambda_{\text{cri}}$, the system undergoes the first order phase transition. At first glance, it seems difficult to find the solution of Eq.(\ref{eq:saddle point}) because $\langle S_H^{}\rangle$ changes discretely at that point. 
 %On the other  hand, when $\phi_2(\lambda_\text{cri}^{})^{4}V_4^{}\geq I_H^{}\geq \phi_1(\lambda_\text{cri}^{})^{4}V_4^{}$, we can not simply find the solution $\lambda^*(I_H^{})$ because the system undergoes the first order transition. 
However, even if $I_H^{}>\phi_1^{}(\lambda_{\text{cri}})^4V_4^{}$, the universe can actually satisfy Eq.(\ref{eq:saddle point}) by keeping the coexisting of two vacua over the whole space:
\be  \langle S_H^{}\rangle=\left(x\phi_1^{}(\lambda_{\text{cri}}^{})+(1-x)\phi_2^{}(\lambda_{\text{cri}}^{})\right)^4\times V_4^{}=I_H^{},
\e
where $x$ represents the ratio of one vacuum to the other vacuum. In other words, the symbol $\langle\cdots\rangle$ also includes the average over the space. See the right panel in Fig.\ref{fig:phase2} for example. This shows a typical distribution of the Higgs vacua when the coexisting is realized. Thus, when $\phi_2^{}(\lambda_{\text{cri}})^4V_4^{}>I_H^{}>\phi_1^{}(\lambda_{\text{cri}})^4V_4^{}$, the system is always the coexisting phase, and $\lambda$ is fixed at $\lambda_{\text{cri}}^{}$. (See again the left panel in Fig.\ref{fig:phase2}.) This result is completely the same as the temperature in statistical mechanics. Finally, when $I_H^{}\geq \phi_2(\lambda_\text{cri}^{})^{4}$, $\lambda$ is fixed to $\lambda^*(I_H^{})$ in the same way as the $I_H^{}\leq \phi_1(\lambda_\text{cri}^{})^{4}$ case. As a result, the system most likely realize the degenerate vacua because the probability of $I_H^{}$ being chosen to be a value in the interval $\phi_2(\lambda_\text{cri}^{})^{4}V_4^{}\geq I_H^{}\geq \phi_1(\lambda_\text{cri}^{})^{4}V_4^{}$ is biggest  \footnote{The numerical value of $\lambda_{\text{cri}}^{}$, of course, depends on the values of the other SM couplings. If the observed Higgs quartic coupling $\lambda\simeq 0.12$ really corresponds to $\lambda_{\text{cri}}$, the top mass must be around $171$GeV which is slightly small compared with the recent analyses: $M_{t}=173.34\pm0.27\pm0.71$GeV \cite{Aaboud:2016igd} and $M_{t}=172.44\pm 0.13\pm0.47$GeV \cite{Khachatryan:2015hba}.  However, the relation between these masses and the pole mass is not yet clear.  
}. This is the derivation of the MPP from the micro-canonical QFT.  %Nature favors the coexisting vacua if it is described by the microscopic picture. 

%This shows a schematic contour of one of the fixed couplings $\lambda$ as a function of $I$. This figure is the counterpart of Fig.\ref{fig:phase1}, and one can see that there is a finite interval $[I^{(1)},I^{(2)}]$ at which two vacua degenerate. %Note that its explicit shape is not important because our discussion here is quite general.
%If we choose the Higgs quartic term $\int d^4x\ \left(H^\dagger H\right)^2$ as $S_i^{}$, $I^{(1)}$ and $I^{(2)}$ are roughly given by
%\be I^{(1)}=V_4^{} \phi_1^{4}(=V_4^{}v_h^{4}),\ I^{(2)}=V_4^{} \phi_2^{4}(\sim V_4^{}M_{pl}^4),
%\e
%where $\phi_1^{}\ (\phi_2^{})$ is the small (large) vacuum expectation value, and the open parentheses represent the SM cases. When $I^{(1)}<I<I^{(2)}$, the solution of Eq.(\ref{eq:saddle point}) is given by the mixture of two vacua:
%\be\left \langle \int d^4x\left(H^\dagger H\right)^2\right\rangle =V_4^{}\times \langle \phi\rangle^4=V_4^{}\times \left(x\phi_1^{}+(1-x)\phi_2^{}\right)^4=I^{}.
%\e
%Thus, if $I$ is firstly chosen to be in $[I^{(1)},I^{(2)}]$, the coupling is dynamically tuned so that two vacua become degenerate. This is the derivation of the MPP from the micro-canonical QFT.
%if $S_i^{}$ is chosen to be one that is related to the Higgs potential. 

Let us now discuss the multi-local theory \cite{MLT}, and show that it predicts the same fine-tuning mechanism as the FNM. For now, it is sufficient to assume that the following partition function is given first:
\be Z^{(\text{MLT})}=\int {}{\cal{D}}\phi\ \exp\left(i\sum_{i} c_i^{}S_i^{}+i\sum_{i,j}c_{ij}^{}S_i^{}S_j^{}+\cdots\right),
\label{eq:multi partition}
\e
where $c_{i}$'s, $c_{i,j}$'s, $\cdots$ are constants. From the microscopic point of view, the multi-locality in Eq.(\ref{eq:multi partition}) comes from physics at the Planck/String scale such as matrix model. By regarding $S_i^{}$'s as variables and introducing the Lagrange multipliers $\lambda_i^{}$'s, we can rewrite Eq.(\ref{eq:multi partition}) as
\aln{ &=\int{\cal{D}}\phi\ \int \prod_i d \lambda _i^{}\ \omega(\lambda_1^{},\lambda_2^{},\cdots)\exp\left(i\sum_{i}\lambda_{i}S_{i}\right)\nonumber
\\
&= \int \left(\prod_i d \lambda _i^{}\right)\ \omega(\overrightarrow{\lambda})\ Z^{(\text{QFT})}\left(\overrightarrow{\lambda}\right),\label{eq:multi partition2}
}
where 
\be \omega(\lambda_1^{},\lambda_2^{},\cdots)=\int\cdots \int \left(\prod_i dS_i^{}\right)\exp\left(i\sum_{i} (c_i^{}-\lambda_i^{})S_i^{}+i\sum_{i,j}c_{ij}^{}S_i^{}S_j^{}+\cdots\right)
\e
 is the Fourier coefficient. One can see that Eq.(\ref{eq:multi partition2}) is the same as Eq.(\ref{eq:mc QFT2}) except for the weight factor $\omega(\overrightarrow{\lambda})$ and $I_i^{}$'s. %\footnote{Although one might think that $I_i^{}$'s do not exists in Eq.(\ref{eq:multi partition2}), they are already included in $S_i^{}$'s in the following sense. Because $S_i^{}$'s in Eq.(\ref{eq:multi partition}) are determined from the microscopic theory such as wormhole theory or matrix model at the Planck/String scale, they are in fact bare quantities. However, because what we want to know is the fine-tuning of the renormalized couplings, we should interpret these bare quantities as $S_i=\tilde{S}_i^{}+I_i^{}$, where $\tilde{S}_i^{}$ is the action with the renormalized coupling.}
The former is not important for the determination of $\lambda_i^{}$'s because it is just a Fourier coefficient of the ordinary function, and does not have a strong dependence on $\lambda_i^{}$'s. Therefore, we can conclude that the couplings are determined by $Z^{(\text{QFT})}(\overrightarrow{\lambda})$ as well as the FNM. In this sense, the FNM and the multi-local theory predict the same fine-tuning mechanism in QFT \footnote{On the other hand, the predicted values of the couplings are generally different each other because $I_i^{}$'s do not exist in the multi-local theory. As for the MPP, however, we can show that it can be derived even in the multi-local theory because the coexisting point of different phases corresponds to a non-analytic point of $\rho(\overrightarrow{\lambda})$. See \cite{MLT} for the details.
}.
 Although we have assumed Eq.(\ref{eq:multi partition}) as the multi-local theory, our conclusion does not change even if we start from more general partition function
\be \tilde{Z}^{(\text{MLT})}=\int{\cal{D}}\phi\ F(S_0^{},S_1^{},\cdots),\label{eq:general partition function}
\e
where $F(S_0^{},S_1^{},\cdots)$ is an ordinary function of $S_i^{}$'s. This is because it can be rewritten like Eq.(\ref{eq:multi partition2}) by doing the Fourier transform. Therefore, the equivalence between the FNM and the multi-local theory is quite general, and it might be interesting to study the explicit form of $F(S_0^{},S_1^{},\cdots)$ in some microscopic theory.

\begin{figure}[t]
\begin{center}
\includegraphics[width=14cm]{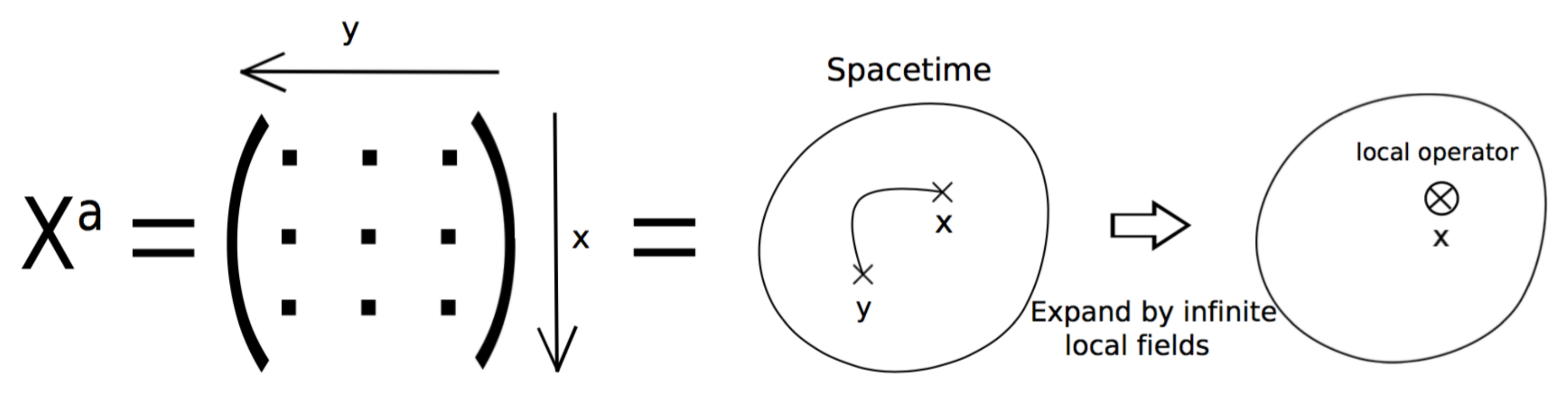}
\end{center}
\caption{The covariant derivative interpretation of matrix model. $i$ and $j$ of a matrix $(X^a)_{ij}$ represent two points on a manifold ${\cal{M}}$ (left), and $(X^a)_{ij}$ operates on a function $f(x)$ as $\sum_{ij}(X^a)_{ij}f(x_i^{})$ (middle). However, by introducing infinite local fields, it can be also represented as a local operator (right).}
\label{fig:matrix}
\end{figure}

Finally, we discuss a concrete example of the multi-local theory from matrix model \cite{Asano:2012mn} for the readers who do not satisfy the above abstract argument. Although matrix model is believed to include gravity and can be a non-perturbative formulation of string, its interpretation is not yet clear \cite{Matrix}. If gravity is really included in matrix model, the diffeomorphism invariance must also be included in the formulation. In the following, we adopt the covariant derivative interpretation \cite{Hanada:2005vr} because it has the manifest diffeomorphism invariance. The action of the IIB matrix model is given by  
\be S_{IIB}^{}=\frac{1}{g^2}\text{Tr}\left(\frac{1}{ 4}[X^a,X^b][X^c,X^d]\eta_{ac}^{}\eta_{bd}^{} +\frac{1}{2}\bar{\Psi}\Gamma^a[X^b,\Psi]\eta_{ab}^{}\right),\label{eq:IIB action}
\e
where $a$, $b$, $c$, and $d$ are the ten dimensional Lorentz indices, $X^a$ and $\Psi$ are a ten dimensional vector and a Majorana spinor respectively, and they are also $N\times N$ hermitian matrices. Eq.(\ref{eq:IIB action}) has the manifest $SO(9,1)$ and $U(N)$ invariances. For now, it is sufficient to consider $X^a$ only. The covariant derivative interpretation interprets $X^a$ as a linear operator acting on smooth functions on a given $D\ (\leq 10)$ dimensional manifold ${\cal{M}}$:
\be X^a\in End(C^\infty({\cal{M}}))=\left\{C^\infty({\cal{M}})\rightarrow C^\infty({\cal{M}})\right\}.\label{eq:derivative interpretation}
\e
Of course, such a interpretation is possible only in the large $N$ limit. %In the following, we take first $D$ components of the ten demential Lorentz index $a$ as the $D$ dimensional Lorentz index $\mu$. Therefore, $(X^0,X^1,\cdots,X^{D-1})$ is a vector, and $(X^D,X^{D+1},\cdots,X^{9})$ are scalars. 
The operation of $X^a$ can be explicitly written as
\aln{ (X^af)(x)&=\int d^D y\ X^a(x,y)f(y)\nonumber
\\
&=\left(S^a(x)+a^{a\mu}(x)\bigtriangledown_\mu^{}+b^{a\mu\nu}(x)\bigtriangledown_\mu^{}\bigtriangledown_\nu^{}+\cdots\right)f(x)\nonumber
\\
&\h{4cm} \text{for } ^\forall f(x)\in C^{\infty}({\cal{M}}),\label{eq:linear operator}
}
where we have expanded the kernel $X^a(x,y)$ by infinite local fields with integer spins and the covariant derivative $\bigtriangledown$ on ${\cal{M}}$. See Fig.\ref{fig:matrix} for example. This shows the intuitive picture of Eq.(\ref{eq:linear operator}). Although $X^a$ naively represents a non-local object, it can be also interpreted as a local operator acting on $C^\infty({\cal{M}})$. Here, note that ten dimensional Lorentz index $a$ has no relation to $D$ dimensional index $\mu$ in general. The classical equation of motion of $X^a$'s is given by
\be \eta_{ab}^{}\left[X^a,\left[X^b,X^c\right]\right]=0.\label{eq:eom}
\e

 %____________________________Matrix__________________________________
%_________________________________________________________________________

\noindent Among the various solutions, the following one 
\be X_a^{}=\begin{cases}i\bigtriangledown_a^{}=e_a^{\mu}(x)\bigtriangledown_\mu^{}&\text{for }a=0,1,\cdots,D-1,
\\ 0&\text{for }a=D,\cdots,9
\end{cases}
\e
is notable because Eq.(\ref{eq:eom}) corresponds to the Einstein equation in this case:
\aln{ \left[\bigtriangledown^a,\left[\bigtriangledown_a^{},\bigtriangledown_b^{}\right]\right]&=\left[\bigtriangledown^a,R_{ab}^{cd}\times\hat{{\cal{O}}}_{cd}^{}\right]\nonumber
\\
&=\left(\bigtriangledown^{a}R_{ab}^{cd}\right)\hat{{\cal{O}}}_{cd}^{}-R_{ab}^{cd}\ \hat{{\cal{O}}}_{cd}^{}\bigtriangledown^{a}=0\nonumber
\\
\leftrightarrow&\bigtriangledown^{a}R_{ab}^{cd}=0,\ R_{ab}=0
}
where $\hat{{\cal{O}}}_{cd}^{}$ is the $D$ dimensional Lorentz generator, and we have interpreted $e_{a}^{\mu}(x)$ as the vielbein field. This fact tells us that gravity can be actually embedded in matrix. Furthermore, we can find the diffeomorphism invariance within the original $U(N)$ symmetry:
\be\delta X^a=i\left[\Lambda,X^a\right],\ \Lambda\in N\times N\ \text{Hermitian matrix}.
\e
The explicit form of $\Lambda$ in the large $N$ limit is given by
\be \Lambda=\lambda(x)+\frac{i}{ 2}\left\{\lambda^\mu(x),\bigtriangledown_\mu^{}\right\}+\frac{i^2}{2}\left\{\lambda^{\mu\nu}(x),\bigtriangledown_{\mu}^{}\bigtriangledown_\nu^{}\right\}+\cdots%\delta^{(D)}(x-y)
\label{eq:explicit infinitesimal un}
\e
as well as $X^a$. Here, we have introduced the anti-commutator $\{,\}$ to make each terms hermitian \footnote{In fact, by taking the hermitian conjugate, we have
\be \Lambda^\dagger=\lambda(x)-\frac{i}{2}\left\{\bigtriangledown_\mu^{\dagger},\lambda^\mu(x)\right\}+\frac{(-i)^2}{2}\left\{\bigtriangledown_{\mu}^{\dagger}\bigtriangledown_\nu^{\dagger},\lambda^{\mu\nu}(x)\right\}+\cdots%\delta^{(D)}(x-y)
,
\e
where $\bigtriangledown_\mu^{\dagger}$ is the derivative which acts on a function on the left (by definition). Thus, by %doing the partial integration, and 
neglecting the total derivative term, we obtain the same expression as Eq.(\ref{eq:explicit infinitesimal un}). Note that we must take the order of $\lambda^{\mu\nu\cdots}(x)$ and $\bigtriangledown_{\mu}^{}$ into consideration to obtain the correct result. 
}. In particular, the second term
\be 
\Lambda=\frac{i}{2}\left\{\lambda^\mu(x),\bigtriangledown_\mu^{}\right\}%\times\delta^{(D)}(x-y)
\e
represents the diffeomorphism of the fields appearing in Eq.(\ref{eq:linear operator}). For example, the scalar $S^a(x)$ transforms as 
\be \delta S^a(x)=i\left[\Lambda,S^{a}\right]=\lambda^\mu(x)\bigtriangledown_\mu^{}S^{a}(x),
\e
and this is actually the diffeomorphism transformation of scalar. Thus, one can see that all the information of a curved manifold ${\cal{M}}$ can be embedded in the matrices $X^a$'s. However, the above argument is not mathematically rigorous because $X_a^{}=e^\mu_a(x)\bigtriangledown_\mu^{}$ is not in fact included in $End(C^{\infty}({\cal{M}}))$ \footnote{$\bigtriangledown_\mu^{}$ is explicitly written as $\bigtriangledown_\mu^{}=\partial_\mu^{}+\omega_{\mu}^{ab}(x)\hat{O}_{ab}$, where $\omega_{\mu}^{ab}(x)$ is a connection, and, as well as $\hat{{\cal{O}}}_{ab}$, its explicit form depends on the representation of a function it operates. For example, when they operate the Lorentz vector, we have $(\hat{O}_{ab})^{\mu}_\nu=\eta^{\mu a}\delta_\nu^{b}-\delta_{\nu}^a\eta^{\mu b}$, $2\omega_{\mu\lambda}^{\nu}(x)=\Gamma_{\mu\lambda}^{\nu}(x)$, where the raising or lowering of indices is done by flat metric. If we consider the product of $X_1^{}=i\bigtriangledown_1^{}:=ie_1^{\mu}\bigtriangledown_\mu^{} $ and $X_2^{}=i\bigtriangledown_2^{}=ie_2^{\mu}\bigtriangledown_\mu^{}$, it is given by 
\be X_1^{}\cdot X_2^{}=(i\bigtriangledown_1^{})(i\bigtriangledown_2^{})=-e_1^{\mu}\partial_\mu^{}(e_2^{\nu}\bigtriangledown_\nu^{} )-e^\mu_1\omega_\mu^{2c}\bigtriangledown_c^{}=-e_1^{\mu}\partial_\mu^{}(e_2^{\nu}\bigtriangledown_\nu^{} )-e^\mu_1\omega_\mu^{2c}X_c^{},
\e
from which we can see that the right hand side is not a product of $X_1^{}$ and $X_2^{}$, but contains other matrices. Therefore, in this naive covariant derivative interpretation, $\bigtriangledown_i^{}$ is not included in $End(C^\infty({\cal{M}}))$.
}. 
To overcome this situation, new formulation of the covariant derivative was proposed in \cite{Hanada:2005vr} where $X^a$'s are interpreted as linear operators acting on smooth functions on the principal bundle $E_{\text{prin}}$ whose base space is ${\cal{M}}$, and fibre is $Spin(D-1,1)$. The result of \cite{Hanada:2005vr} is mathematically rigorous, so we can actually realize a curved spacetime by matrix. For our present purpose, however, we do not need such a rigorous result, and we take the above naive covariant derivative interpretation in the following discussion.%to know the origin of multi-locality. So, in the following, we will not write the $Spin(D-1,1)$ element explicitly.

%
 %____________________________ propagator__________________________________
\begin{figure}
\begin{center}
\includegraphics[width=12cm]{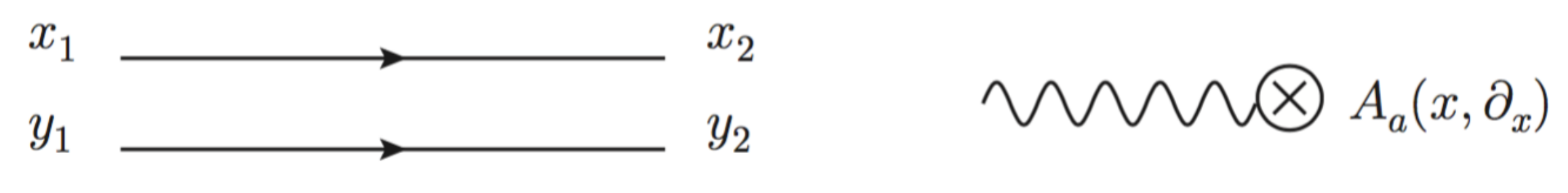}
\end{center}
\caption{Left: The propagator of the fluctuation $\phi(x,y)$. Right: The background field $A_a^{}(x,\partial_x^{})$ or $A_a^{}(y,\partial_y^{})$.}
\label{fig:propagator}
\end{figure}
%_________________________________________________________________________
%

To examine the effective action, we use the background field method:
\be X^a(x,y)=\tilde{X}^a(x,y)+\phi^a(x,y),
\e
where $\tilde{X}^a(x,y)$ is the background field, and $\phi^a(x,y)$ represents the fluctuation which should be integrated out. The bosonic part of the action Eq.(\ref{eq:IIB action}) now becomes
\aln{ S_{IIB}^{}\bigg|_{boson}=\frac{1}{4g^2}\text{Tr}\bigg([\tilde{X}^a,\tilde{X}^b][\tilde{X}_a^{},\tilde{X}_b^{}]&+4[\tilde{X}^a,\tilde{X}^b][\tilde{X}_a^{},\phi_b^{}]+2[\tilde{X}^a,\phi^b][\tilde{X}_a^{},\phi_b^{}]+2[\tilde{X}^a,\tilde{X}^b][\phi_a^{},\phi_b^{}]\nonumber
\\
&+2[\tilde{X}^a,\phi^b][\phi_a^{},\tilde{X}_b^{}]+4[\phi^a,\tilde{X}^b][\phi_a,\phi_b^{}]+[\phi^a,\phi^b][\phi_a^{},\phi_b^{}]\bigg),\label{eq:boson part}
}
where the second term can be always eliminated by the field redefinition of $\phi^a$. Although there are three quadratic terms %of $\phi^a$'s 
in Eq.(\ref{eq:boson part}), it is sufficient to consider the third term in Eq.(\ref{eq:boson part}) for our qualitative understanding:
\be S_{\phi^2}^{}=\frac{1}{2}\text{Tr}\left([\tilde{X}^a,\phi][\tilde{X}^{a},\phi]\right),\label{eq:simple action}
\e
where we have put $g=1$, and picked up one component of $\phi^a$'s for simplicity. The effective theory on the flat spacetime can be obtained by expanding the background field around the flat derivative: 
%Furthermore, by taking the so far argument into consideration, it is natural to expand the background field around the flat derivative:
\be \tilde{X}^a(x,y)=\delta^{(D)}(x-y)\times\begin{cases}i\partial^a+A^a(x,\partial_x^{})&\text{for }a=0,1,\cdots,D-1
\\
A^a(x,\partial_x^{})&\text{for }a=D,\cdots,9
\end{cases},
\e
where $A^a(x,\partial_x^{})$ is a function of $x$ and its derivative, and contains infinite local fields like Eq.(\ref{eq:linear operator}) and Eq.(\ref{eq:explicit infinitesimal un}). In this case, Eq.(\ref{eq:simple action}) becomes
\aln{ S_{\phi^2}^{}=\frac{1}{2}\int d^Dx\int d^Dy&\left[\left(\frac{\partial}{\partial x^\mu}+\frac{\partial}{\partial y^\mu}-i{\cal{A}}_\mu^{}(x,\partial_x^{};y,\partial_y^{})\right)\phi(x,y)\right]^2\nonumber
\\
&+\frac{1}{2}\int d^Dx\int d^Dy\left[{\cal{A}}_a^{}(x,\partial_x^{};y,\partial_y^{})\phi(x,y)\right]^2,\label{eq:tree action}
}
where 
\be {\cal{A}}_a^{}(x,\partial_x^{};y,\partial_y^{})=A_a^{}(x,\partial_x^{})-A_a^{}(y,\partial_y^{}).\label{eq:background}
\e
From Eq.(\ref{eq:tree action}), we can read the propagator of $\phi(x,y)$ as
\aln{ D(x_1^{},y_1^{};x_2^{},y_2^{})&=G((x_1^{}+y_1^{})-(x_2^{}+y_2^{}))\times \delta^{(D)}((x_1^{}-y_1^{})-(x_2^{}-y_2^{})),\label{eq:propagator}
%\\
%&=G(\tilde{x}_1^{}-\tilde{x}_2^{})\times \delta^{(D)}(\xi_1^{}-\xi_2^{}),
}
where $G(x)$ is the propagator of free massless scalar. We represent this propagator and the three (four) point vertex between $\phi$ and the background $A_a^{}(x,\partial_x^{})$ or $A_a^{}(y,\partial_y^{})$ by the double lines and the circled cross mark respectively. See Fig.\ref{fig:propagator} for example. Note that only one external wavy line sticks to the vertex because $\phi(x,y)$ interacts $A_a^{}(x,\partial_{x}^{})$ or $A_a^{}(y,\partial_{y}^{})$.

\begin{figure}[!t]
\begin{center}
\includegraphics[width=12cm]{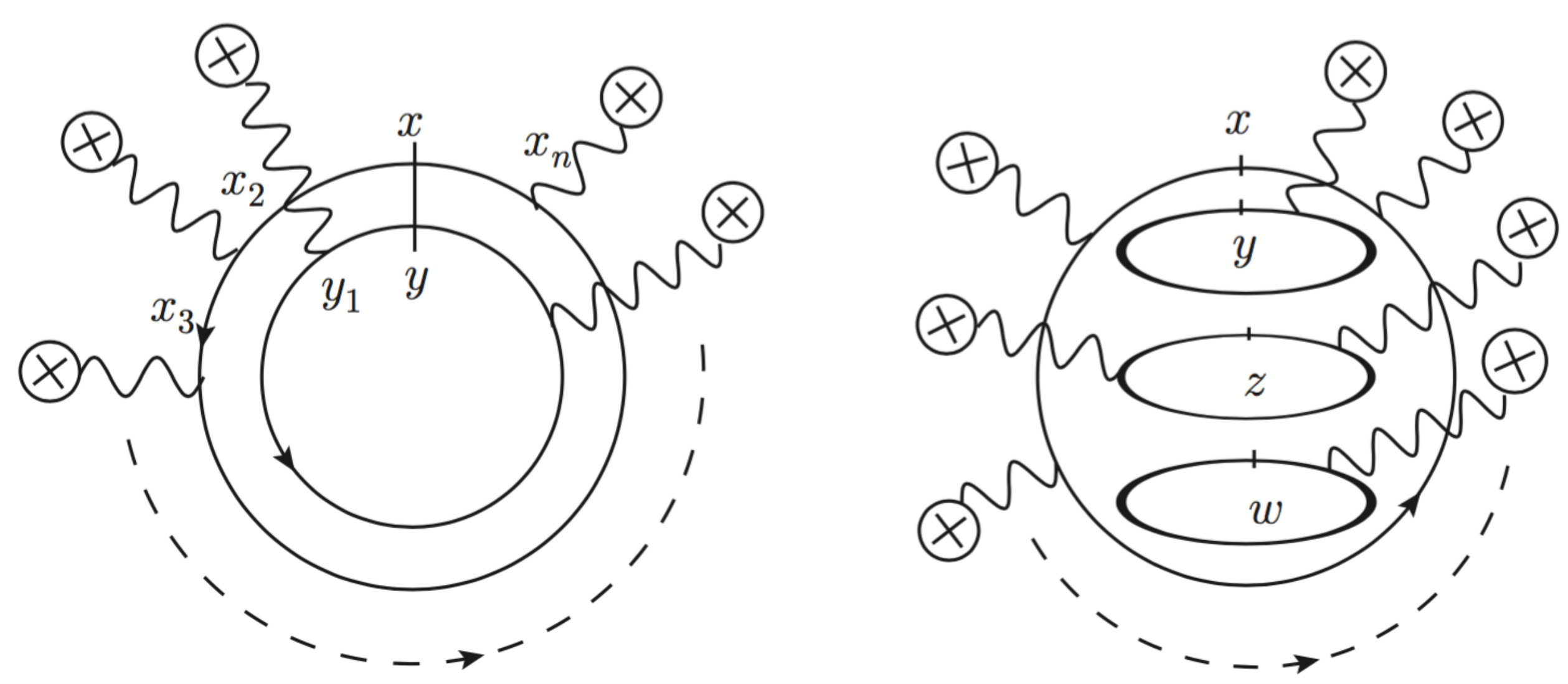}
\end{center}
\caption{Typical two-loop (left) and four-loop (right) diagrams in the covariant derivative interpretation of matrix model. The number of loops corresponds to the number of the products of the effective actions.
}
\label{fig:effective}
\end{figure}

We can now calculate the effective action
\be \exp\left(iS_{\text{eff}}[{\cal{A}}]\right)=\int{\cal{D}}\phi\ \exp\left(iS_{\phi^2}^{}\right)\label{eq:effective action}
\e
based on the loop expansion. This is absolutely the same calculation as the effective potential in QFT 
\footnote{For example, if we neglect the quartic interaction $\phi\phi {\cal{A}}_a^2$ in Eq.(\ref{eq:tree action}), $S_{\text{eff}}[{\cal{A}}]$ can be understood as the generating functional of $n$ point correlation function of $\phi$. Thus,  it can be expanded as
\aln{ S_{\text{eff}}[{\cal{A}}]=\sum_{n=0}^{\infty}\int d^Dx_1^{}\int d^Dy_1^{}\cdots\int d^Dx_n^{}\int d^Dy_n^{}&G^{(n)}(x_1^{},y_1^{};x_2^{},y_2^{};\cdots;x_n^{},y_n^{})\nonumber
\\
&\partial^\mu{\cal{A}}_\mu(x_1^{},\partial_{x_1^{}}^{};y_1^{},\partial_{y_{1}^{}}^{})\times \cdots\times\partial^\mu{\cal{A}}_\mu(x_n^{},\partial_{x_n^{}}^{};y_n^{},\partial_{y_{n}^{}}^{}),
}
where $G^{(n)}$ is the $n$ point correlation function. %and we have considered three point vertex for simplicity. 
The one-loop effective action corresponds to considering the one-loop diagram of $G^{(n)}$.
}. For example, at two-loop level\footnote{Here, we use the term ``$n$-loop'' as the number of the spacetime integrals. Thus, one-loop of $\phi$ corresponds to two-loop in our case.
}, we must calculate the one-loop closed diagram of $\phi(x,y)$ with arbitrary $n$ insertions of $A_a^{}$:
\aln{ &\int d^Dx\int d^Dy\prod_{i=1}^{n}\Bigg[\int d^Dx_i^{}\int d^Dy_i^{}D(x_{i+1}^{},y_{i+1}^{};x_{i}^{},y_{i}^{})\nonumber
\\
 &\h{2cm}\times\begin{cases} \partial^\mu A_\mu(x_i^{},\partial_{x_i^{}}^{}),\ \partial^\mu A_\mu(y_i^{},\partial_{y_i^{}}^{})\\
\h{1.5cm} \text{or}\\
 A_a(x_i^{},\partial_{x_i^{}}^{})^2,\ A_a(y_i^{},\partial_{y_i^{}}^{})^2,\  2A^a(x_i^{},\partial_{x_i^{}}^{})A_a(y_i^{},\partial_{y_i^{}}^{})
\end{cases}\Bigg]\times D(x_{1}^{},y_{1}^{};x,y)
%\\
%&= \int d^Dx\int d^Dy\left(\prod_{i=1}^{n}\int d^Dx_i^{}\tilde{A}_i^{(x)}(x_i^{},\partial_{x_i^{}})\right)\left(\prod_{i=1}^{n}\int d^Dy_i^{}\tilde{A}_i^{(y)}(y_i^{},\partial_{y_i^{}})\right) 
\label{eq:1 loop}
}
where $x_{n+1}^{}=x$, $y_{n+1}^{}=y$. %and $\tilde{\partial}^\mu$ represents the derivative with respect to the center coordinate $x^\mu+y^\mu$.
See the left figure in Fig.\ref{fig:effective} for example. For our present purpose, it is sufficient to consider the cubic interaction $\phi\phi\partial^\mu A_\mu^{}$ because our conclusion does not change even if we consider the quartic interaction $\phi\phi A_a^2$. In order to obtain the effective action written by local fields, let us expand each $A_a^{}(x_i^{},\partial_{x_i^{}}^{})$'s like Eq.(\ref{eq:explicit infinitesimal un}):
\aln{ A^{a}(x_i^{},\partial_{x_i^{}}^{})&=\hat{A}^a(x_i^{})+\frac{i}{2}\left\{\hat{A}^{a\mu}(x_i^{}),\partial_\mu^{(i)}\right\}+\frac{i^2}{2}\left\{\hat{A}^{a\mu\nu}(x_i^{}),\partial_{\mu}^{(i)}\partial_\nu^{(i)}\right\}+\cdots\nonumber
\\
&=\sum_{k=0}^{\infty}\frac{i^n}{k!}\{\hat{A}^{a\mu_1^{}\cdots\mu_k^{}}(x_i^{}),\partial_{\mu_1^{}}^{(i)}\cdots\partial_{\mu_k^{}}^{(i)}\}\nonumber
\\
&=\sum_{k=0}^{\infty}\frac{i^n}{k!}\sum_{m=0}^{\infty}\frac{1}{m!}\hat{A}^{a\mu_1^{}\cdots\mu_k^{}}_{\nu_1^{}\cdots\nu_m^{}}(x)\{\tilde{x}_i^{\nu_1^{}}\cdots \tilde{x}_i^{\nu_m^{}},\tilde{\partial}_{\mu_1^{}}^{(i)}\cdots\tilde{\partial}_{\mu_k^{}}^{(i)}\}\nonumber
\\
&:=\sum_{k=0}^{\infty}\frac{i^n}{k!}\sum_{m=0}^{\infty}\frac{1}{m!}\hat{A}^{a\{\mu\}}_{\{\nu\}}(x)\hat{\cal{P}}^{(i)\{\nu\}}_{\{\mu\}}(\tilde{x}_i^{},\tilde{\partial}^{(i)}).\label{eq:local vertex}
}
where $\tilde{x}_i^{}=x_i^{}-x$, $\tilde{\partial}^{(i)}_\mu$ is the derivative with respect to $\tilde{x}_i^{\mu}$, and we have expanded $\hat{A}^{a\mu_1^{}\cdots\mu_k^{}}(x_i^{})$ around $x$ from the second line to the third line. Note that we have also used 
\be \frac{\partial}{\partial x_i^{\mu}}=\frac{\partial}{\partial \tilde{x}_i^{\mu}}.
\e
By substituting Eq.(\ref{eq:local vertex}) into Eq.(\ref{eq:1 loop}), we generally obtain
\aln{ &\int d^Dx \left(\hat{A}^{\alpha_1^{}\{\mu_1^{}\}}_{\{\nu_1^{}\}}(x)\cdot \cdots \cdot\hat{A}^{\alpha_l^{}\{\mu_l^{}\}}_{\{\nu_l^{}\}}(x)\right) \int d^Dy \left( \hat{A}^{\beta_1^{}\{\mu_1^{}\}}_{\{\nu_1^{}\}}(y)\cdot \cdots \cdot\hat{A}^{\beta_f^{}\{\mu_f^{}\}}_{\{\nu_f^{}\}}(y)\right)\times\uwave{\left(\prod_{j=1}^n\int d^D\tilde{x}_j^{}\int d^D\tilde{y}_j^{}\right)}\nonumber
\\
&\uwave{\times D(x,y;\tilde{x}_{n}^{},\tilde{y}_{n}^{})\times\begin{cases}\tilde{\partial}^{(n)}_{\alpha}\hat{\cal{P}}^{(n)\{\nu\}}_{\{\mu\}}(\tilde{x}_n^{},\tilde{\partial}^{(n)})
\\
\quad\quad\text{ or }
\\
\tilde{\partial}^{(n)}_{\beta}\hat{\cal{P}}^{(n)\{\nu\}}_{\{\mu\}}(\tilde{y}_n^{},\tilde{\partial}^{(n)})
\end{cases}
\times D(\tilde{x}_{n}^{},\tilde{y}_n^{};\tilde{x}_{n-1}^{},\tilde{y}_{n-1}^{})\times\cdots\times \begin{cases}\tilde{\partial}^{(1)}_{\alpha}\hat{\cal{P}}^{(1)\{\nu\}}_{\{\mu\}}(\tilde{x}_1^{},\tilde{\partial}^{(1)})
\\
\quad\quad\text{ or }
\\
\tilde{\partial}^{(1)}_{\beta}\hat{\cal{P}}^{(1)\{\nu\}}_{\{\mu\}}(\tilde{y}_1^{},\tilde{\partial}^{(1)})
\end{cases}}\nonumber
\\
&\uwave{\times D(\tilde{x}_{1}^{},\tilde{y}_1^{};x,y)},\label{eq:1 loop 2}
}
where the numbers $l$ and $f$ depend on the choice of vertexes, and we have changed the variable of each of the integrations from $x_i^{}\ (y_i^{})$ to $\tilde{x}_i^{}\ (\tilde{y}_i^{})$. Note that we have not explicitly written the lower index of $\alpha$ ($\beta$) in $\tilde{\partial}_\alpha^{(i)}$ ($\tilde{\partial}_\beta^{(i)}$) because there are many possible ways of their contractions. %$\alpha^i$ of $\hat{A}^{\alpha_i^{}}$ can contract with any $\partial_\alpha^{}\hat{{\cal{P}}}$.
Furthermore, from Eq.(\ref{eq:propagator}), we can see that $D(\tilde{x}_1^{},\tilde{y}_1^{};x,y)$ and $D(x,y;\tilde{x}_n^{},\tilde{y}_n^{})$ do not depend on $x$ and $y$ \footnote{In fact, we have
\aln{ D(\tilde{x}_1^{},\tilde{y}_1^{};x,y) &= G(\tilde{x}_1^{}+\tilde{y}_1^{}+x+y-(x+y))\times\delta^{(D)}\left((\tilde{x}_1^{}-\tilde{y}_1^{}+x-y)-(x-y)\right)\nonumber
\\
&=G(\tilde{x}_1^{}+\tilde{y}_1^{})\times\delta^{(D)}\left(\tilde{x}_1^{}-\tilde{y}_1^{}\right).\nonumber
}
}. 
As a result, the underlined part in Eq.(\ref{eq:1 loop 2}) is just a coefficient, and we finally obtain the factorized bi-local action:
\be c_{\alpha_1^{}\cdots \alpha_l^{}\beta_1^{}\cdots\beta_f^{}\{\mu_1^{}\}\cdots\{\mu_l^{}\},\{\mu_1^{}\}\cdots\{\mu_f^{}\}}^{\{\nu_1^{}\}\cdots\{\nu_l^{}\},\{\nu_1^{}\}\cdots\{\nu_f^{}\}}\int d^Dx \left(\hat{A}^{\alpha_1^{}\{\mu_1^{}\}}_{\{\nu_1^{}\}}(x)\cdot \cdots \cdot\hat{A}^{\alpha_l^{}\{\mu_l^{}\}}_{\{\nu_l^{}\}}(x)\right) \int d^Dy \left( \hat{A}^{\beta_1^{}\{\mu_1^{}\}}_{\{\nu_1^{}\}}(y)\cdot \cdots \cdot\hat{A}^{\beta_f^{}\{\mu_f^{}\}}_{\{\nu_f^{}\}}(y)\right).
\e
The reason why we have obtained the bi-local action originates in the number of loops in a diagram: Because one-loop of $\phi$ corresponds to two-loop of the spacetime integrals, the corresponding effective action becomes bi-local. This feature descends to higher loop diagrams. When we consider $m$-loop diagram with arbitrary number of insertion of $A_a^{}$'s, we generally obtain
\be \sum_{i_1^{},\cdots,i_m^{}}c_{i_1^{}\cdots i_m^{}}\int d^Dx_1^{}{\cal{O}}_{i_1^{}}(x_1^{})\times \int d^Dx_2^{}{\cal{O}}_{i_2^{}}(x_2^{})\times\cdots\times \int d^Dx_m^{}{\cal{O}}_{i_m^{}}(x_m^{}),
\e
where $i_k^{}$'s represent the various indexes. See the right figure of Fig.\ref{fig:effective} for example. Here we show the typical four-loop closed diagram where the self interaction of $\phi$ comes from the last term in Eq.(\ref{eq:tree action}). Thus, the effective action $S_{\text{eff}}[{\cal{A}}]$ in the covariant derivative interpretation of matrix model generally contains many multi-local actions, and the effective couplings become dynamical from the previous argument.

 %____________________________ effective action__________________________________
%_________________________________________________________________________

%%%%%%%%%%%%%%%%%%%% ACKNOWLEDGMENTS %%%%%%%%%%%%%%%%%%%%
\section*{Acknowledgement} 
This work is supported by the Grant-in-Aid for Japan Society for the Promotion of Science (JSPS) Fellows No.27$\cdot$1771.

\appendix 
\def\thesection{Appendix \Alph{section}}


\begin{thebibliography}{unsrt}

\bibitem{MPP1} 
  %\cite{Froggatt:1995rt}
  C.~D.~Froggatt and H.~B.~Nielsen,
  ``Standard model criticality prediction: Top mass 173 +- 5-GeV and Higgs mass 135 +- 9-GeV,''
  Phys.\ Lett.\ B {\bf 368}, 96 (1996)
  doi:10.1016/0370-2693(95)01480-2
  [hep-ph/9511371];
  %%CITATION = doi:10.1016/0370-2693(95)01480-2;%%
  %167 citations counted in INSPIRE as of 01 May 2016
  %
  %\cite{Froggatt:2001pa}
  C.~D.~Froggatt, H.~B.~Nielsen and Y.~Takanishi,
  ``Standard model Higgs boson mass from borderline metastability of the vacuum,''
  Phys.\ Rev.\ D {\bf 64}, 113014 (2001)
  doi:10.1103/PhysRevD.64.113014
  [hep-ph/0104161];
  %%CITATION = doi:10.1103/PhysRevD.64.113014;%%
  %59 citations counted in INSPIRE as of 01 May 2016
  %
  %\cite{Froggatt:2004bh}
  C.~D.~Froggatt, H.~B.~Nielsen and L.~V.~Laperashvili,
  ``Hierarchy-problem and a bound state of 6 t and 6 anti-t,''
  Int.\ J.\ Mod.\ Phys.\ A {\bf 20} (2005) 1268
  doi:10.1142/S0217751X0502416X
  [hep-ph/0406110];
  %%CITATION = doi:10.1142/S0217751X0502416X;%%
  %43 citations counted in INSPIRE as of 07 Aug 2016
  %
  %\cite{Nielsen:2012pu}
  H.~B.~Nielsen,
  ``PREdicted the Higgs Mass,''
  Bled Workshops Phys.\  {\bf 13}, no. 2, 94 (2012)
  [arXiv:1212.5716 [hep-ph]].
  %%CITATION = ARXIV:1212.5716;%%
  %24 citations counted in INSPIRE as of 12 Aug 2016
  
  \bibitem{MPP2} 
  %\cite{Buttazzo:2013uya}
  D.~Buttazzo, G.~Degrassi, P.~P.~Giardino, G.~F.~Giudice, F.~Sala, A.~Salvio and A.~Strumia,
  ``Investigating the near-criticality of the Higgs boson,''
  JHEP {\bf 1312}, 089 (2013)
  doi:10.1007/JHEP12(2013)089
  [arXiv:1307.3536 [hep-ph]];
  %%CITATION = doi:10.1007/JHEP12(2013)089;%%
  %510 citations counted in INSPIRE as of 01 Sep 2016
  %\cite{Iso:2012jn}
  S.~Iso and Y.~Orikasa,
  ``TeV Scale B-L model with a flat Higgs potential at the Planck scale - in view of the hierarchy problem -,''
  PTEP {\bf 2013}, 023B08 (2013)
  doi:10.1093/ptep/pts099
  [arXiv:1210.2848 [hep-ph]];
  %%CITATION = doi:10.1093/ptep/pts099;%%
  %92 citations counted in INSPIRE as of 01 May 2016
   %\cite{Kawana:2014zxa}
  K.~Kawana,
  ``Multiple Point Principle of the Standard Model with Scalar Singlet Dark Matter and Right Handed Neutrinos,''
  PTEP {\bf 2015}, 023B04 (2015)
  doi:10.1093/ptep/ptv006
  [arXiv:1411.2097 [hep-ph]];
  %%CITATION = doi:10.1093/ptep/ptv006;%%
  %16 citations counted in INSPIRE as of 01 May 2016
  %\cite{Hamada:2014wna}
  Y.~Hamada, H.~Kawai, K.~y.~Oda and S.~C.~Park,
  ``Higgs inflation from Standard Model criticality,''
  Phys.\ Rev.\ D {\bf 91}, 053008 (2015)
  doi:10.1103/PhysRevD.91.053008
  [arXiv:1408.4864 [hep-ph]];
  %%CITATION = doi:10.1103/PhysRevD.91.053008;%%
  %36 citations counted in INSPIRE as of 01 Sep 2016
  %
  %\cite{Kawana:2015tka}
  K.~Kawana,
  %``Criticality and inflation of the gauged B \UTF{2013} L model,''
  PTEP {\bf 2015}, 073B04 (2015)
  doi:10.1093/ptep/ptv093
  [arXiv:1501.04482 [hep-ph]];
  %%CITATION = doi:10.1093/ptep/ptv093;%%
  %20 citations counted in INSPIRE as of 12 Aug 2016
  %\cite{Hamada:2015ria}
  Y.~Hamada, H.~Kawai and K.~y.~Oda,
  ``Eternal Higgs inflation and the cosmological constant problem,''
  Phys.\ Rev.\ D {\bf 92}, 045009 (2015)
  doi:10.1103/PhysRevD.92.045009
  [arXiv:1501.04455 [hep-ph]];
  %%CITATION = doi:10.1103/PhysRevD.92.045009;%%
  %19 citations counted in INSPIRE as of 01 Sep 2016
  %
  %\cite{Hamada:2015fma}
  Y.~Hamada and K.~Kawana,
  ``Vanishing Higgs Potential in Minimal Dark Matter Models,''
  Phys.\ Lett.\ B {\bf 751}, 164 (2015)
  doi:10.1016/j.physletb.2015.10.006
  [arXiv:1506.06553 [hep-ph]].
  %%CITATION = doi:10.1016/j.physletb.2015.10.006;%%
  %6 citations counted in INSPIRE as of 01 May 2016
%\cite{Haba:2015rha}
  N.~Haba and Y.~Yamaguchi,
  ``Vacuum stability in the $U(1)_\chi$ extended model with vanishing scalar potential at the Planck scale,''
  PTEP {\bf 2015}, no. 9, 093B05 (2015)
  doi:10.1093/ptep/ptv121
  [arXiv:1504.05669 [hep-ph]];
  %%CITATION = doi:10.1093/ptep/ptv121;%%
  %11 citations counted in INSPIRE as of 01 May 2016
  %
  %\cite{Haba:2015nwl}
  N.~Haba, H.~Ishida, R.~Takahashi and Y.~Yamaguchi,
  ``Gauge coupling unification in a classically scale invariant model,''
  JHEP {\bf 1602}, 058 (2016)
  doi:10.1007/JHEP02(2016)058
  [arXiv:1511.02107 [hep-ph]];
  %%CITATION = doi:10.1007/JHEP02(2016)058;%%
  %1 citations counted in INSPIRE as of 01 May 2016
  %
  %\cite{Haba:2015qbz}
  N.~Haba, H.~Ishida, N.~Kitazawa and Y.~Yamaguchi,
  ``A new dynamics of electroweak symmetry breaking with classically scale invariance,''
  Phys.\ Lett.\ B {\bf 755}, 439 (2016)
  doi:10.1016/j.physletb.2016.02.052
  [arXiv:1512.05061 [hep-ph]];
  %%CITATION = doi:10.1016/j.physletb.2016.02.052;%%
  %2 citations counted in INSPIRE as of 01 May 2016
  %\cite{Haba:2016gqx}
  N.~Haba, H.~Ishida, N.~Okada and Y.~Yamaguchi,
  ``Multiple-point principle with a scalar singlet extension of the Standard Model,''
  arXiv:1608.00087 [hep-ph].
  %%CITATION = ARXIV:1608.00087;%%

%\cite{Hamada:2015dja}
\bibitem{Hamada:2015dja} 
  Y.~Hamada, H.~Kawai and K.~Kawana,
  ``Natural solution to the naturalness problem: The universe does fine-tuning,''
  PTEP {\bf 2015}, no. 12, 123B03 (2015)
  doi:10.1093/ptep/ptv168
  [arXiv:1509.05955 [hep-th]].
  %%CITATION = doi:10.1093/ptep/ptv168;%%
  %5 citations counted in INSPIRE as of 31 Aug 2016
  
  %\cite{Coleman:1988tj}
 \bibitem{Coleman:1988tj} 
  S.~R.~Coleman,
  ``Why There Is Nothing Rather Than Something: A Theory of the Cosmological Constant,''
  Nucl.\ Phys.\ B {\bf 310}, 643 (1988).
  doi:10.1016/0550-3213(88)90097-1.
  %%CITATION = doi:10.1016/0550-3213(88)90097-1;%%
  %699 citations counted in INSPIRE as of 26 Apr 2016

%\cite{Asano:2012mn}
\bibitem{Asano:2012mn} 
  Y.~Asano, H.~Kawai and A.~Tsuchiya,
  ``Factorization of the Effective Action in the IIB Matrix Model,''
  Int.\ J.\ Mod.\ Phys.\ A {\bf 27}, 1250089 (2012)
  doi:10.1142/S0217751X12500893
  [arXiv:1205.1468 [hep-th]].
  %%CITATION = doi:10.1142/S0217751X12500893;%%
  %4 citations counted in INSPIRE as of 19 Aug 2016

 %\cite{Aaboud:2016igd}
\bibitem{Aaboud:2016igd} 
  M.~Aaboud {\it et al.} [ATLAS Collaboration],
  ``Measurement of the top quark mass in the $t\bar{t}\to$ dilepton channel from $\sqrt{s}=8$ TeV ATLAS data,''
  arXiv:1606.02179 [hep-ex].
  %%CITATION = ARXIV:1606.02179;%%
  %7 citations counted in INSPIRE as of 31 Aug 2016

%\cite{Khachatryan:2015hba}
\bibitem{Khachatryan:2015hba} 
  V.~Khachatryan {\it et al.} [CMS Collaboration],
  ``Measurement of the top quark mass using proton-proton data at ${\sqrt{(s)}}$ = 7 and 8 TeV,''
  Phys.\ Rev.\ D {\bf 93}, no. 7, 072004 (2016)
  doi:10.1103/PhysRevD.93.072004
  [arXiv:1509.04044 [hep-ex]].
  %%CITATION = doi:10.1103/PhysRevD.93.072004;%%
  %44 citations counted in INSPIRE as of 31 Aug 2016

 \bibitem{MLT} 
  %\cite{Kawai:2011qb}
  H.~Kawai and T.~Okada,
  ``Solving the Naturalness Problem by Baby Universes in the Lorentzian Multiverse,''
  Prog.\ Theor.\ Phys.\  {\bf 127}, 689 (2012)
  doi:10.1143/PTP.127.689
  [arXiv:1110.2303 [hep-th]];
  %%CITATION = doi:10.1143/PTP.127.689;%%
  %25 citations counted in INSPIRE as of 01 May 2016
  %
  %\cite{Hamada:2014ofa}
  Y.~Hamada, H.~Kawai and K.~Kawana,
  ``Evidence of the Big Fix,''
  Int.\ J.\ Mod.\ Phys.\ A {\bf 29}, 1450099 (2014)
  doi:10.1142/S0217751X14500997
  [arXiv:1405.1310 [hep-ph]];
  %%CITATION = doi:10.1142/S0217751X14500997;%%
  %15 citations counted in INSPIRE as of 01 May 2016
  %
  %\cite{Hamada:2014xra}
  Y.~Hamada, H.~Kawai and K.~Kawana,
  ``Weak Scale From the Maximum Entropy Principle,''
  PTEP {\bf 2015}, 033B06 (2015)
  doi:10.1093/ptep/ptv011
  [arXiv:1409.6508 [hep-ph]];
  %%CITATION = doi:10.1093/ptep/ptv011;%%
  %13 citations counted in INSPIRE as of 01 May 2016
  %
  %\cite{Kawana:2016oof}
  K.~Kawana,
  ``Classical conformality in the Standard Model from Coleman's theory,''
  arXiv:1605.00436 [hep-ph].
  %%CITATION = ARXIV:1605.00436;%%
  %1 citations counted in INSPIRE as of 12 Aug 2016
  
  
  %\cite{Ishibashi:1996xs}
\bibitem{Matrix} 
  N.~Ishibashi, H.~Kawai, Y.~Kitazawa and A.~Tsuchiya,
  ``A Large N reduced model as superstring,''
  Nucl.\ Phys.\ B {\bf 498}, 467 (1997)
  doi:10.1016/S0550-3213(97)00290-3
  [hep-th/9612115];
  %%CITATION = doi:10.1016/S0550-3213(97)00290-3;%%
  %861 citations counted in INSPIRE as of 30 Aug 2016
  %\cite{Aoki:1999vr}
  H.~Aoki, N.~Ishibashi, S.~Iso, H.~Kawai, Y.~Kitazawa and T.~Tada,
  ``Noncommutative Yang-Mills in IIB matrix model,''
  Nucl.\ Phys.\ B {\bf 565}, 176 (2000)
  doi:10.1016/S0550-3213(99)00633-1
  [hep-th/9908141];
  %%CITATION = doi:10.1016/S0550-3213(99)00633-1;%%
  %281 citations counted in INSPIRE as of 30 Aug 2016
  %\cite{Steinacker:2010rh}
  H.~Steinacker,
  ``Emergent Geometry and Gravity from Matrix Models: an Introduction,''
  Class.\ Quant.\ Grav.\  {\bf 27}, 133001 (2010)
  doi:10.1088/0264-9381/27/13/133001
  [arXiv:1003.4134 [hep-th]].
  %%CITATION = doi:10.1088/0264-9381/27/13/133001;%%
  %95 citations counted in INSPIRE as of 30 Aug 2016

%\cite{Hanada:2005vr}
\bibitem{Hanada:2005vr} 
  M.~Hanada, H.~Kawai and Y.~Kimura,
  ``Describing curved spaces by matrices,''
  Prog.\ Theor.\ Phys.\  {\bf 114}, 1295 (2006)
  doi:10.1143/PTP.114.1295
  [hep-th/0508211];
  %%CITATION = doi:10.1143/PTP.114.1295;%%
  %53 citations counted in INSPIRE as of 16 Aug 2016
  %\cite{Kawai:2007zz}
  H.~Kawai,
  ``Curved space-times in matrix models,''
  Prog.\ Theor.\ Phys.\ Suppl.\  {\bf 171}, 99 (2007).
  doi:10.1143/PTPS.171.99.
  %%CITATION = doi:10.1143/PTPS.171.99;%%
  %2 citations counted in INSPIRE as of 16 Aug 2016
 

\end{thebibliography}
\end{document}